\documentclass{aa}
\usepackage{aabib99}
\usepackage{psfig,amssym}
\newcommand{\etal}{{et al.}}
\begin{document}
   \thesaurus{03  (11.01.2;
                   11.02.1;
                   11.10.1;
                   11.14.1;
                   13.18.1)}

\title{Flat radio-spectrum galaxies and BL\,Lacs: {\bf Part I:} core properties}
\titlerunning{Flat radio-spectrum galaxies and BL\,Lacs: I}
\author{J. Dennett-Thorpe\inst{1,2} \and M. J. March{\~a}\inst{2}}
\institute{Kapteyn Institute, Postbus 800, 9700 AV Groningen, Netherlands
     \and Observat{\'o}rio Astron{\'o}mico de Lisboa, Tapada da Ajuda, 1300 Lisbon, Portugal}
\offprints{J. Dennett-Thorpe}
\mail{jdt@astro.rug.nl}

\date{Received /Accepted}
\maketitle

\begin{abstract}

This paper concerns the relationship of BL\,Lacs and flat--spectrum
weak emission--line galaxies. We compare the weak emission--line
galaxies and the BL\,Lacs in a sample of 57 flat--spectrum objects
(March{\~a} et al. 1996), using high-frequency radio and non-thermal
optical flux densities, spectral indices and polarization
properties. We consider whether objects which are not `traditional'
BL\,Lacs---due to their larger emission line strengths, and larger
Ca\,II spectral breaks---are simply star\-light diluted BL\,Lacs.  Their
broad-band spectral properties are consistent with this
interpretation, but their radio polarization may indicate more subtle
effects. Comparison of the weak emission--line galaxies and the BL\,Lacs
shows that, on average, the former have steeper spectra between 8 and
43\,GHz, and are less polarized at 8.4\,GHz. This is consistent with
many of the weak--lined objects being at larger angles to the line of
sight than the BL\,Lacs.  In addition to this population, we indicate
a number of the weak emission--line galaxies which may be `hidden
BL\,Lacs': relativistically boosted objects very close to the line of
sight with an apparently weak AGN.

\end{abstract}


\section{Introduction}

Explanations of different source properties as a manifestation of
orientation dependent effects have been successful in both studies of
high--powered radio galaxies and quasars, and in radio--weak
Seyferts. Obscuration of the nuclear regions by a putative dusty disk
at large angles to the line of sight is believed to turn Type I
sources (Seyfert Is and quasars) into Type 2 sources (Seyfert 2s and
narrow-line radio galaxies) (Scheuer 1987; Barthel 1989; Antonucci
1993).  Viewing--angle dependent effects attributed to relativistic
bulk flows (superluminal motion, rapid variability, core dominance,
one-sided appearance) are also seen in the powerful radio sources
\cite[and references therein]{urr95}
\nocite{sch87-contrib,bar89,ant93}

The effects of orientation at intermediate radio luminosities remain
much less clear (see Urry \& Padovani 1995 for a review). It is now
well accepted that BL\,Lacs are predominantly FRI radio sources seen at
close angles to the line of sight. Many of their distinctive
characteristics can be attributed to relativistic motion close to the
line of sight: the FRI/BL\,Lac inner jets are believed to have $\gamma$
between 2 and 20 \cite{urr95,lar97}, and BL\,Lacs are thought to be
those objects seen within $\sim 30^\circ$ to the line of sight
\cite{urr95,ghi93}. The role of obscuration at these intermediate
luminosities however remains uncertain (see e.g. Falcke, Gopal-Krishna
\& Biermann, 1995; Jackson \& March{\~a}, 1999; Chiaberge, Capetti \&
Celotti, 1999).  Furthermore, many important questions remain
unanswered or disputed: the relationship between the BL\,Lacs selected
by different means (are the X-ray selected BL\,Lacs closer to the line
of sight than the radio selected ones, or vice versa? or are they
objects with underlying different energy
distributions? \cite{fos97,pad95,mar86}); the frequency of BL\,Lacs which have
FRII parents \cite{kol92,sta97,jac99a}. \nocite{jac99,fal95b,chi99}

In this paper we analyse the radio properties of a sample of
flat--spectrum radio sources \cite{mjm96}. Such a population should
contain a high fraction of sources at small angles to the line of
sight.  Indeed, March{\~a} \etal~ found a high fraction of BL\,Lacs and
candidate BL\,Lacs, as well as both weak and strong emission--line
objects. The canonical parent population of BL\,Lacs, the FRIs, are
generally weak emission--line objects and so the weak emission--line
objects in the sample could be objects which were somewhat further
from our line of sight. (The relationship, if any, to the strong
emission--line objects is less clear.)

Browne \& March{\~a} (1993) suggested that objects which do not meet
the generally used criteria for classification of an object as a BL\,Lac
(EW$<$5\AA, CaII break contrast $\leq$0.25; Stickel et al. 1991;
Stocke et al. 1991) may still be BL\,Lacs, but whose identifying
characteristics are diluted by starlight.\nocite{bro93} Therefore new
criteria were adopted by March{\~a} et al. (1996) [hereafter MBIS] to
allow for this possible starlight dilution. Under these criteria,
objects were classed as BL\,Lac candidates if they had CaII break
contrasts lower than 0.4 and EWs that could be up to $\sim$ 30 \AA. As
an extension of this idea, it is possible that some of the
weak emission--line galaxies are not actually (substantially) further
from the line of sight than the BL\,Lacs, but that they are `hidden
BL\,Lacs'. That is, their non-thermal emission at optical wavelengths
was fainter than the identified BL\,Lacs, and therefore their nuclear
emission is completely swamped by the starlight of the host galaxy. In
this scenario the jet is still relativistic and beamed towards the
observer at the same, or slightly \\greater, angles as occurs in BL\,Lacs.

In this paper we address:

\noindent (i) whether the BL\,Lac `candidates' found by MBIS have broad
band spectral properties consistent with their being bona-fide, but
starlight diluted, BL\,Lacs.

\noindent (ii) the relationship of the BL\,Lacs with the weak emission
line galaxies (WLRG), by comparison of their broad band spectral
properties and core radio polarization.  We analyse whether any of the
WLRG are candidates for `hidden' BL\,Lacs.

In a future paper, we use the results of this study and the radio
morphologies to further analyse the relationship of the BL\,Lacs with
the WLRG and, further, to investigate relationship of the broad emission--line
galaxies (BLRG).

Sect. \ref{sect:obs} records the observational parameters, Sect.
\ref{sect:anal} discusses the methods used in analysis of the data and
presents some of the measured source parameters in tabular
form. Sect.~\ref{sect:results} considers these in the light of the
spectral classification of the sources, the discussion of which is
deferred to Sect.~\ref{sect:discussion}.


\section{The sample}

The sample is the 57 objects of MBIS. These objects satisfy the
selection criteria S$_{\rm 5GHz} >$ 200mJy, $\delta > 20^\circ$ and
m$_v <$17. The Hubble relation for radio sources shows that these
criteria will enable selection of an essentially complete
redshift--limited sub-sample with z$<$0.1. Thus MBIS define a complete
sample of 33 flat--spectrum objects with S$_{\rm 5GHz} >$ 200mJy at
z$<$0.1.  MBIS divide their sample into 6 categories: the known
BL\,Lacs and new BL\,Lacs discovered by them which satisfy the Stocke
et al. (1991) equivalent width--CaII break contrast criteria
(EW$<$5\AA, contrast $<$0.25), the `candidate BL\,Lacs', the weak
emission--line galaxies, the strong emission--line objects (both
broad-- and narrow--lined), and two `hybrid' objects which have
properties intermediate between BL\,Lacs and quasars
\cite{jac99}. There are also a few objects for which they were unable
to obtain spectra and whose classification is therefore unknown.

We assign the class `BL\,Lac' to those objects (both the `known' and
`new' BL\,Lacs of MBIS) which fall within the Stocke et
al. criteria. Those which fall outside this region, but within the
extended criteria of MBIS, we classify as `candidate BL\,Lacs'
(although due to a change in definition of the contrast our
classification of `candidate' BL\,Lacs differs slightly from that used
in MBIS -- see Sect. \ref{sec:opt}). Thus Mrk\,501 (B\,1652+398) is classed
as a `candidate' BL\,Lac, on account of its EW(H$\alpha$) (MBIS).
This object is already often classed as a BL\,Lac due to its extreme
optical brightness, which makes it a subject of frequent study.

  In this paper we are primarily interested in the categories of
BL\,Lacs, candidate BL\,Lacs and weak emission line objects, although we
will comment on the status of the strong emission line objects and
hybrids, and for completeness include information on the objects of
unknown spectral type.

\section{Observations and data reduction}
\label{sect:obs}

The observations were taken with the NRAO VLA CnD array on 29
September and  5, 18 and 27 October 1997. 43\,GHz receivers are only
available on 13 antennae: we therefore observed 22\,GHz and 43\,GHz in one
subarray, and 8 and 15\,GHz simultaneously in the other
subarray. On-line integration times were 3.3s.  Observing parameters
and typical rms noise levels are shown in Table~\ref{tab:obs}. Fast
switching was used at 43\,GHz where this was beneficial (the existence
of a nearby bright good calibrator at this frequency), but in general
this was not done: the phase stability in the relatively compact array
was sufficient, and moreover the relatively strong, point-like
structure of the sources themselves allowed for most sources to be
self-calibrated. Absolute flux density and polarization position angle
calibration were done with reference to 3C\,286. Due to failure of the
telescope, three sources have no data at any frequency
(B\,1646+499, B\,1652+398 \& B\,1703+223), and one only at 8\,GHz (B\,1658+302).

\begin{table}
\caption{The observational parameters}
\begin{tabular}{llllll}
\hline
\multicolumn{5}{c}{Oct 1997. VLA CnD array}\\
\hline
\multicolumn{2}{c}{frequencies} &$\Delta\nu$& beam &ave. \# & rms\\
\multicolumn{2}{c}{GHz} & MHz &      ('')    &visib.& mJy/bm\\
\hline
8.435 & 8.485 & 50 & 8&1200 & 0.25 \\
14.965&14.915 & 50 & 3&2100 & 0.4 \\     
22.485&22.435 & 50 & 3&6500 & 0.6 \\    
43.315&43.365 & 50 & 1&3200 & 2 \\
\hline
\end{tabular}
\label{tab:obs}
\end{table}

\section{Data analysis}
\label{sect:anal}
\subsection{The high frequency radio observations}

The flux densities were measured using the {\sc aips} image and
(u,v)--plane fitting tasks IMFIT and UVFIT.  At these resolutions, all
the sources are dominated by a single component at 8\,GHz and
higher. It was therefore found that a good fit was obtained to all
sources with a single Gaussian and a zero-level offset with slope, if
the fit was restricted to the region $\sim$ 3 times FWHM of the
restoring beam. The values from IMFIT obtained in this manner were
used as the flux densities of the core. The error was estimated using
the difference between this and those obtained by UVFIT.  At 8\,GHz
the `fitting' error calculated in this manner is very small. At
43\,GHz the UVFIT and IMFIT tasks sometimes yield different results
due to the very low signal-to-noise ratio on any given visibility. The
total quoted error also includes a 3\% calibration error.

\begin{figure}
\psfig{file=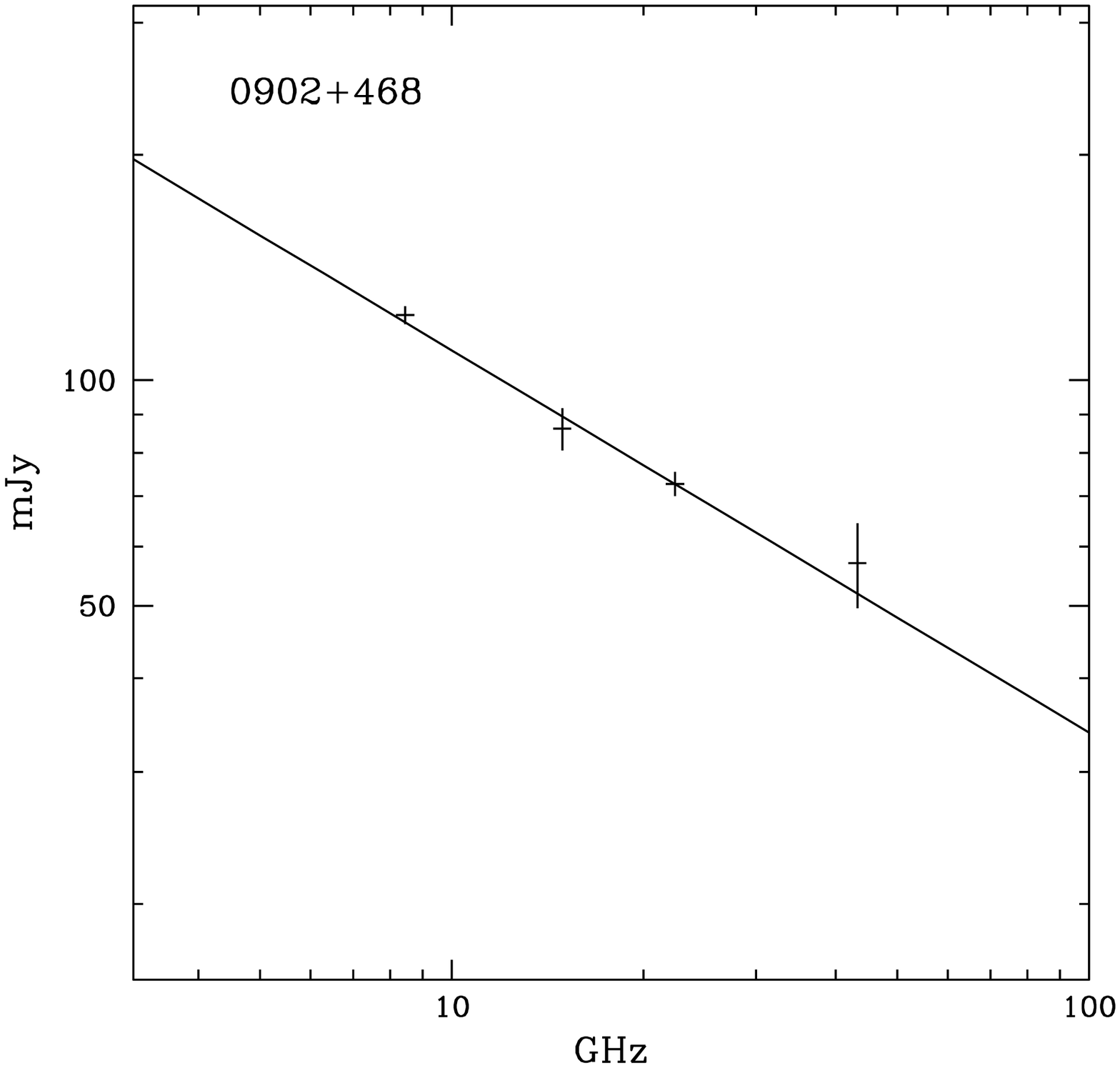,height=8cm}
\psfig{file=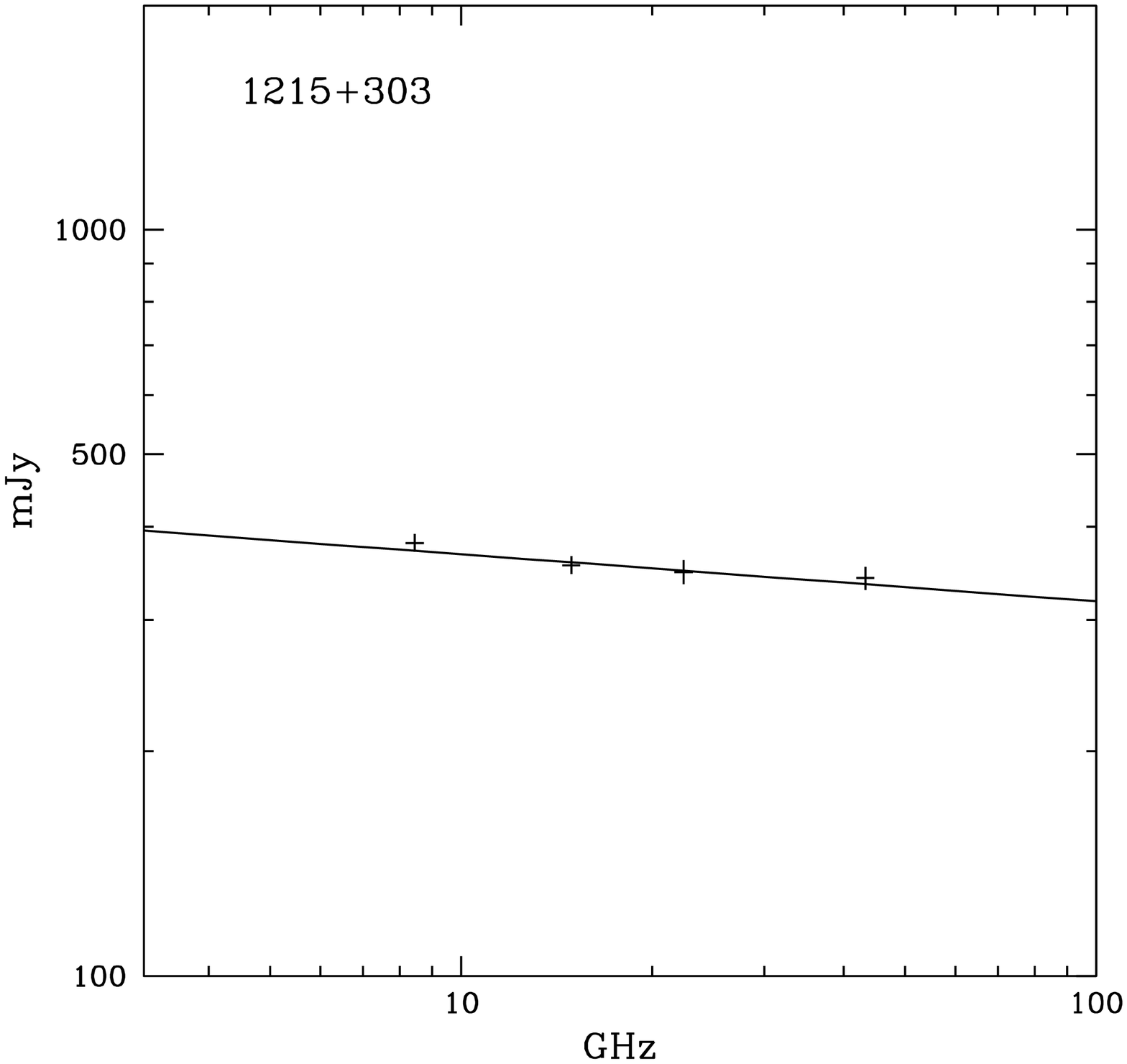,height=8cm}
\caption{Examples of the spectra and fits obtained: a falling spectrum
WLRG, B0902+468, and BL\,Lac B1215+303 with a flat spectrum. }
\label{fig:radio.ps}
\end{figure}

We fitted a power--law spectrum to all the high frequency data
(S$\propto \nu^{-\alpha}$), using a $\chi^2$ fit, taking account of
the errors on the derived flux densities. Only one source (B\,1404+286, a
known GPS) had a (down-curving) spectra clearly inconsistent with the
straight line fit. B\,1404+286 was included in the sample of MBIS as
it is a serendipitously flat spectrum source between 1.4 \& 5\,GHz, as
its spectral peak falls near these frequencies.  The error on the spectral
fits is simply the weighted $\chi^2$. Examples of the spectra and the
fits are shown in fig~\ref{fig:radio.ps}.
The core polarized flux density was calculated from similar fits to
Ricean corrected polarized intensity maps. The 8.4\,GHz flux
densities, spectral indices between 8.4 and 43\,GHz and percentage
polarization measured at 8.4\,GHz, with accompanying errors, are
presented in columns 3 to 8 of Table~\ref{tab:res}, which presents
the observational results for all sources.

\subsection{The non-thermal contribution to the optical flux densities}
\label{sec:opt}

 The observed optical continuum is the sum of the contribution
 from the host galaxy ($S_{\rm o}^{\rm G}$) and that arising from the
 AGN ($S_{\rm o}^{\rm AGN}$). As the objective is to establish the
 broad--band properties of a sample of AGNs, it is important that the
 AGN contribution alone is considered.

 The optical spectrum of early--type galaxies shows a prominent flux
depression shortward of 4000\AA~ which is referred to as the Ca\,II
break or the 4000\AA~ break. The strength of this break is quantified
by the break contrast, C:

\begin{equation}
C = \frac{S_{\nu}^{+}-S_{\nu}^{-}}{S_{\nu}^{+}} . 
\label{defc}
\end{equation} 

\noindent where $S_\nu^{-,+}$ are the flux densities at wavelengths respectively
longer and shorter than the Ca\,II break.

Dressler \& Shectman (1987) studied a large sample of galaxies and
found that ellipticals had a mean break contrast of $<$C$^{\rm G}>$ =
0.49. Inspection of their data showed that less than 5\% of early-type
galaxies have break contrasts below 0.4. It is therefore a reasonable
assumption that an elliptical galaxy with C $\leq$ 0.4 has an extra
source of continuum which reduces the strength of the Ca\,II break. In
other words, if the measured break contrast is found to be below 0.4,
there is good statistical evidence to assume that it is due to an
extra source of continuum, in particular, a non-thermal source due to
the AGN. We therefore calculate the non-thermal contribution from the
measured break contrast and total optical flux, according to:
\nocite{dre87a}

\begin{equation}
S_{\rm o}^{\rm AGN} = \frac{<C^{\rm G}> - C}{<C^{\rm G}>} S_{\rm o} . 
\label{nthermal}
\end{equation}

There is no reason why objects showing break contrasts larger
than 0.4 should not hold extra sources of continuum, but these will
not be detected via the measurement of the contrast (i.e. they are
the `hidden BL\,Lacs'). 

It should be noted that the above equation is valid only for sources
in which the line contribution to the optical spectrum is negligible,
and where C $\leq$ 0.4. For those sources where there is either a
significant emission line contribution or C$>$ 0.4, the estimated
optical non-thermal contribution is likely to be
overestimated. Furthermore, any non-thermal optical contribution for
those sources with C$>$0.5 cannot be estimated in this manner.  In
these latter cases, we assigned a break contrast of 0.49, and obtained
an upper limit of $S_{\rm o}^{\rm AGN}$. Finally, we mention that the
S$_{\rm o}$ should be measured close to the rest-frame wavelength of the
Ca\,II break (4000\,\AA). We measured the flux density at a
rest--frame wavelength of 5500\,\AA, in order to be consistent with
the wavelength usually considered when determining the broad--band
spectral indices.  We will show that our results are not dependent on
this choice.

 We use Eq.~\ref{nthermal} to estimate the AGN contribution to the
optical flux for the sources for which there is a break contrast given
in MBIS. For sources without measured break contrasts (the BL\,Lacs),
either the spectra of MBIS were used to determine the non-thermal
emission, or where these were not taken (due to the status of an
object as a known BL\,Lac), the V band magnitudes from the literature
were used to estimate the non-thermal optical flux density (assuming
$\alpha_{\rm opt}$ =0). These were: B\,0109+224 \cite{pus80},\\ B\,1147+245,
B\,1215+303 and B\,1219+285 \cite{tap76}. A spectrum from the
literature was used for B\,2116+81 \cite{sti93}.

The values of the contrast and optical flux densities are tabulated in
Table~\ref{tab:res}. The contrast values given here are slightly
different from those in MBIS since the latter were calculated using
S$_{\lambda}$ in Eq.~\ref{defc}, instead of the definition used
here (S$_{\nu}$). The new values (consistent with Dressler \& Shectman (1987))
result in contrasts $>$0.4 for the sources B\,0149+710, B\,1144+352,
B\,1551+239, and \\B\,1703+223. In what follows we no longer classify
these sources as `candidate' BL\,Lacs as in MBIS, but comment on them
individually where this is thought appropriate.  We note however, that
this leaves only candidate objects very close to the Stocke et al
criteria in EW-contrast space, and B\,1744+260 (which lies closer to
the `hybrids' B\,0125+487 and B\,1646+499.) The new classifications
are in column 2 of Table~\ref{tab:res}.

\begin{table*}
\caption{The source properties: 8$\,$--$\,$43$\,$GHz and 5000\AA}
\label{tab:res}
\begin{tabular}{lllrrrrrrrl}
\hline
Source  & type&contrast&
\multicolumn{2}{c}{S(8GHz)}&\multicolumn{2}{c}{$\alpha$(8.4--43GHz)}
&\%P(8GHz)& S$_{\rm o}$(5500\AA) &S$_{\rm o}^{\rm AGN}$(5500\AA)  &other\\ 
       & &     & (mJy)  &+/-&        &+/-&        & (mJy) &(mJy)&name\\
\hline
B\,0035+227  & WE& 0.57 &    181&   5&    0.88& 0.06& $<$0.81&  0.30 &$<$0.01 \\
B\,0046+316  & NE& 0.35 &    260&   8&    0.38& 0.05&    2.22&  1.32 &   0.40 & Mrk348\\
B\,0055+300  & WE& 0.56 &    798&  24&    0.06& 0.04& $<$0.18&  3.96 &$<$0.08 &NGC315\\
B\,0109+224  & KB& 0    &   1472&  44&   -0.46& 0.04&    1.62&  2.0  &   2.0 \\
B\,0116+319  & WE& 0.56 &   1107&  33&    0.75& 0.05&    0.22&  0.60 &$<$0.01 &4C31.04\\
B\,0125+487  & HY& 0.38 &    177&   5&   -0.15& 0.05&    3.48&  0.31 &   0.07 \\
B\,0149+710  & WE& 0.46 &    457&  14&    0.03& 0.03&    2.67&  0.63 &   0.05 \\
B\,0210+515  & KB& 0.26 &    182&   6&    0.28& 0.04&    3.29&  0.93 &   0.44 \\
B\,0251+393  & BE& 0.00 &    456&  14&    0.13& 0.04&    1.67&  0.30 &   0.30 \\
B\,0309+411  & BE& 0.25 &    458&  14&   -0.04& 0.05&    0.92&  0.22 &   0.11 \\
B\,0316+413  & BE& --   &  22531& 677&    0.41& 0.03&    0.25&   --  &   --   &3C84/PerA\\
B\,0321+340  & NE& 0.00 &    493&  15&    0.38& 0.04&    4.53&  0.97 &   0.97 \\
B\,0651+410  & WE& 0.55 &    288&   9&    0.17& 0.04& $<$0.43&  2.01 &$<$0.04 \\
B\,0651+428  & CB& 0.26 &    156&   5&    0.25& 0.04&    1.31&  0.36 &   0.18 \\
B\,0716+714  & KB& 0.00 &    719&  22&    0.01& 0.06&    5.23&123.00 & 123.00 \\
B\,0729+562  & WE& 0.61 &    123&   4&    0.75& 0.07& $<$1.02&  1.00 &$<$0.02 \\
B\,0733+597  & WE& 0.58 &    185&   6&    0.34& 0.06& $<$0.62&  1.62 &$<$0.03 \\
B\,0806+350  & KB& 0.25 &    108&   3&    0.44& 0.06&    2.98&  0.66 &   0.32 \\
B\,0848+686  & WE& 0.54 &     44&   3&    0.13& 0.16&    2.54&  2.53 &$<$0.05 \\
B\,0902+468  & WE& 0.54 &    122&   4&    0.51& 0.05& $<$0.99&  0.71 &$<$0.01 \\
B\,0912+297  & KB& 0.07 &    178&   7&    0.39& 0.03&    2.69&  1.58 &   1.36 \\
B\,1055+567  & CB& 0.08 &    196&   6&   -0.06& 0.03&    1.31&  3.26 &   2.73 \\
B\,1101+384  & KB& 0.06 &    656&  20&    0.17& 0.03&    2.37& 60.90 &  54.20 &Mrk421\\
B\,1123+203  & CB& 0.11 &    428&  14&    0.24& 0.03& $<$0.28&  0.86 &   0.68 \\
B\,1133+704  & CB& --   &    147&   5&   -0.01& 0.03&    3.27&   --  &   --   &Mrk180\\
B\,1144+352  & WE& 0.51 &    372&  11&    0.29& 0.03&    0.44&  1.70 &$<$0.03 \\
B\,1146+596  & WE& 0.56 &    416&  12&    0.33& 0.05& $<$0.31&  7.33 &$<$0.15 \\
B\,1147+245  & KB& 0   &     770&  27&    0.19& 0.03&    1.38&  1.51 &   1.51 \\
B\,1215+303  & KB& 0   &     380&  11&    0.06& 0.03&    4.16&  1.64 &   1.64 &ON235\\
B\,1217+295  & WE& --  &     132&   4&    0.65& 0.07&    0.65&   --  &   --   &NGC4278\\
B\,1219+285  & KB& 0   &     696&  21&    0.23& 0.05&    3.64&  1.02 &   1.02 &WCom\\
B\,1241+735  & WE& 0.49 &    139&   5&    0.15& 0.03&    8.02&  1.25 &   0.01 \\
B\,1245+676  & WE& 0.56 &    136&   4&    0.56& 0.05& $<$1.02&  0.56 &$<$0.01 \\
B\,1254+571  & BE& --   &    245&   7&    0.79& 0.04& $<$0.50&   --  &   --   &Mrk231\\
B\,1404+286  & BE& 0.25 &   1932&  59&0.99$\star$& 0.05& 0.11&  1.72 &   0.86 &OQ208\\
B\,1418+546  & KB& 0.18 &    616&  18&   -0.07& 0.03&    3.01&  2.06 &   1.33 &OQ530\\
B\,1421+511  & BE& 0.00 &    132&   4&    0.01& 0.04&    1.06&  0.51 &   0.51 \\
B\,1424+240  & KB& 0.00 &    254&   8&    0.20& 0.04&    2.63&  4.28 &   4.28 \\
B\,1532+236  & NE& --   &    148&   5&    1.03& 0.12& $<$0.84&   --  &   --   &Arp220\\
B\,1551+239  & WE& 0.47 &    132&   5&   -0.07& 0.05& $<$0.88&  0.46 &   0.03 \\
B\,1558+595  & WE& 0.57 &    109&   4&    0.89& 0.13& $<$1.06&  2.03 &$<$0.04 \\
B\,1645+292  & KB& 0.34 &     74&   3&    0.58& 0.15&    4.68&  0.35 &   0.12 \\
B\,1646+499  & HY& 0.37 &     --&  --&      --& --  &  --    &  0.76 &   0.20 \\
B\,1652+398  & CB& 0.18 &     --&  --&      --& --  &  --    & 11.80 &   7.47 &Mrk501\\
B\,1658+302  & WE& 0.54 &     86&   2&      --& --  & $<$1.97&  1.18 &$<$0.02 \\
B\,1703+223  & WE& 0.53 &     --&  --&      --& --  &  --    &  0.81 &$<$0.02 \\
B\,1744+260  & CB& 0.36 &    266&   8&    0.08& 0.04&    1.55&  0.22 &   0.06 \\
B\,1755+626  & WE& 0.59 &    150&   5&    0.17& 0.03&    1.62&  2.35 &$<$0.05 &NGC6521\\
B\,1807+698  & KB& 0.15 &   1501&  45&   -0.07& 0.04&    3.11&  4.37 &3.09    &3C371\\  
B\,1959+650  & KB& 0.10 &    254&   8&   -0.12& 0.03&    1.61& 1.45 &   1.16  \\
B\,2116+818  & BE& --   &    135&   6&   -0.35& 0.04& $<$1.09& 0.96 &   0.96  \\
B\,2202+363  & WE& 0.55 &    121&   4&   -0.12& 0.03& $<$1.26& 0.72 &$<$0.01  \\
B\,2214+201  & UK&  --  &    109&   8&    0.59& 0.11&    9.54&  --  &   --    \\
B\,2217+259  & BE& 0.35 &    208&   6&   -0.08& 0.03& $<$0.80& 0.52 &   0.15  \\
B\,2319+317  & UK&  --  &    623&  19&    0.06& 0.03&    0.19&  --  &   --    \\
B\,2320+203  & WE& 0.53 &    151&   6&    0.21& 0.04&    1.41& 1.59 &$<$0.03  \\
B\,2337+268  & UK&  --  &    100&   6&    0.20& 0.05&    2.89&  --  &   --    &NGC7728\\
\hline                                                                 
\end{tabular}

Spectral types (outlined in Sect.~\ref{sec:opt}): KB and CB are
known and candidate BL\,Lacs respectively. BE and NE are strong
emission line objects with broad and narrow lines respectively. WE are
objects with weak emission lines and a spectrum dominated by stellar
light. HY are hybrids, and UK are of unknown spectral type.\\ $\star$
down-curving spectrum not well fit by straight line.
\end{table*}

\begin{figure}
\psfig{file=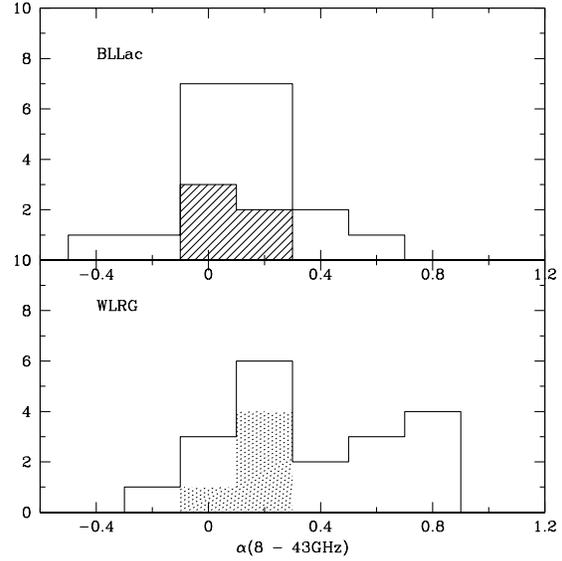,height=8cm}
\caption{Histograms of the high frequency spectral indices of (top
panel) the BL\,Lacs and candidates; candidate BL\,Lacs hatched and (bottom
panel) weak emission line galaxies. The shaded region in the bottom
panel corresponds to potentially `hidden' BL\,Lacs (see Sect.~\ref{sect:hidd}.)}
\label{fig:hi-rad.ps}
\end{figure}

\begin{figure}
\psfig{file=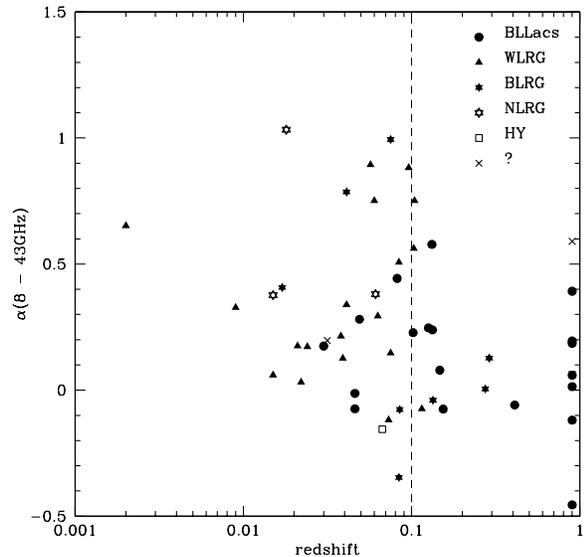,height=8cm}
\caption{High frequency radio spectral index as a function of redshift
for the different classes. The dotted line indicates the redshift
limit corresponding to the `complete sample' of MBIS. Objects of
unknown redshift have been placed arbitrarily at $z$=1}
\label{fig:with-z}
\end{figure}

\section{Results}
\label{sect:results}

\begin{table*}
\caption{Mean properties of classes of objects }
\begin{tabular}{lrrrr}
\hline
category           & BL\,Lac & CBL\,Lac & WLRG & BLRG \\
\hline
$S_{\rm 8GHz}(mJy)$        & 561$\pm$461   &239$\pm$116    &272$\pm$269     &2950$\pm$7364\\
$\alpha^8_{43}$        & 0.13$\pm$0.27 & 0.10$\pm$0.14 & 0.35$\pm$0.32  & 0.22$\pm$0.43\\
$\alpha^{\rm radio}_{\rm opt*}$& 0.47$\pm$0.13 &0.58$\pm$0.15  & --             & -- \\
\%P (8GHz)             &3.0$\pm$1.1    & 1.5$\pm$1.0   & $<$1.4$\pm$1.7 & 0.8$\pm$0.5\\
\hline
\end{tabular}

The mean percentage polarization of the WLRG at 8.4\,GHz is calculated using the upper limits
\label{tab:spect}
\end{table*}

\subsection{High frequency radio spectra}

Firstly we consider high frequency radio spectra of the BL\,Lacs and the
`candidate' BL\,Lacs (S$\propto \nu^{-\alpha}$). As can be seen on the
top panel of Fig.~\ref{fig:hi-rad.ps}, the candidate BL\,Lacs (hatched)
are indistinguishable from the BL\,Lacs (white) by their high frequency
radio spectra. (A Kolmogorov--Smirnov (K-S) test rejects the
hypothesis that they are drawn from the same population only at the
13\% level.) This is in marked contrast to the weak emission--line
galaxies: a comparison of the top and bottom panels shows that the
WLRG have a steeper radio spectra than the BL\,Lacs. This is confirmed
by a K-S test: the hypothesis that the WLRG and BL\,Lacs (including
candidate BL\,Lacs) are drawn from the same population is rejected at
the 88\% level.

 As the BL\,Lacs are generally at higher redshifts than the WLRG
(although there are 7 BL\,Lacs with unknown redshifts), the intrinsic
rest--frame difference in the spectra of the two classes is likely, if
anything, to be greater than the difference observed. However,
Fig.~\ref{fig:with-z} shows there is no dependence of the spectral
index $\alpha^{8}_{43}$ on redshift for any of the different classes.

The BLRG and the BL\,Lacs have a similar distributions of
$\alpha^8_{43}$. The one hybrid which was successfully observed,
B\,0125+487, has a rising spectrum in this frequency range.

\begin{figure}
\psfig{file=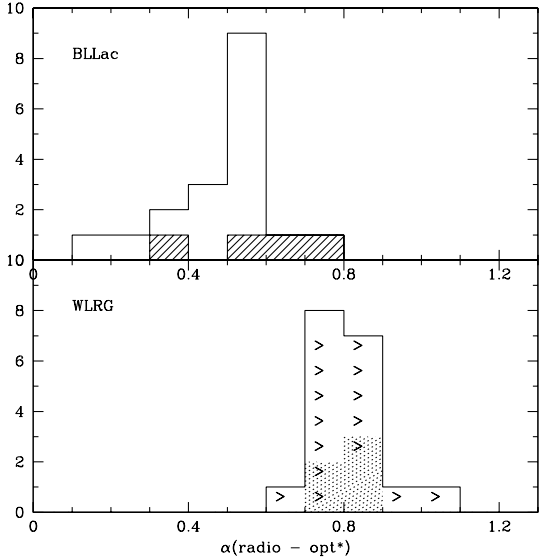,height=8cm}
\caption{Histograms of the radio to non-thermal optical spectral indices of (top
panel) the BL\,Lacs and candidates; candidate BL\,Lacs hatched and (bottom
panel) weak emission line galaxies. The shaded region as for Fig.~\ref{fig:hi-rad.ps}.}
\label{fig:histopt.ps}
\end{figure}

\begin{figure}
\psfig{file=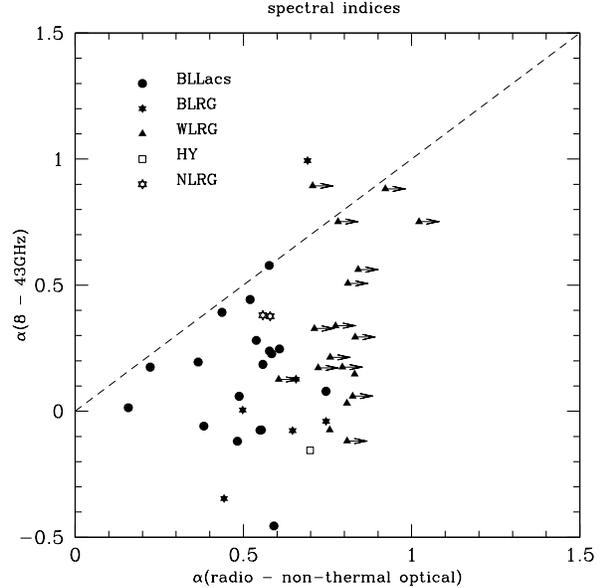,height=8cm}
\caption{Non-thermal spectral indices. The optical flux densities used
are only the non-thermal contribution at 5500\AA, as explained in the text. }
\label{fig:rad-opt.ps}
\end{figure}

\subsection{Radio to optical spectra}

The radio--optical non-thermal spectral index can be used to assess
the similarity of the `candidate' BL\,Lacs and the known BL\,Lacs over a
large frequency range. The top panel of Fig.~\ref{fig:histopt.ps}
shows the radio--optical spectra of the BL\,Lacs and (hatched)
candidates.  For the known BL\,Lacs the mean spectral index
$\bar{\alpha}_{\rm opt*}^{\rm rad} = 0.47\pm0.13$, and for the candidates
$\bar{\alpha}_{\rm opt*}^{\rm rad} = 0.58\pm0.15$, where opt$*$ refers to the
non-thermal contribution to the optical flux.  That they are drawn
from the same population is rejected by a K-S test with a probability of
88\%.  A second estimate of the non-thermal optical flux densities
from S. Anton (priv. comm.), combined with our data calculated from
literature values, gives spectral indices of 0.50$\pm$0.14 for the
BL\,Lacs and 0.63$\pm$0.17 for the candidates.  This second estimate is
not calculated at 5500\AA~ in the rest-frame, but over wavelengths
selected to be free of emission lines nearer the 4000\AA~ break. Using
these values, the same parent population is rejected at the 95\%
level.

Thus we see that the spectra of the candidate BL\,Lacs show some marginal
indication of steepening more rapidly than those of the known BL\,Lacs
in the sample. The candidates and known BL\,Lacs do not show
significantly different distributions in 8.4\,GHz or optical (5500\AA)
flux densities: either the total or the AGN component.

The radio--optical non-thermal spectral index is less useful for
 comparison of the BL\,Lacs with the WLRG because the measurement of the
 AGN component at 5500\AA~ is necessarily dominated by the error in
 the uncertainty of the underlying stellar spectrum. However a clear
 difference between the populations can be seen in
 figs.~\ref{fig:histopt.ps} \& \ref{fig:rad-opt.ps}. In these plots
 the points marked with lower limits refer to a calculation of the
 non-thermal contribution as outlined in Sect.~\ref{sec:opt}. Using
 these upper limits as measurements, we can reject the possibility
 that the samples are the same at greater than the 99.99\% level.

The BLRG and the BL\,Lacs show similar distributions of non-thermal
optical to radio spectral indices, $\alpha_{\rm opt*}^{\rm rad}$, illustrated
in Fig.~\ref{fig:rad-opt.ps}.

\subsection{Radio polarization}

Our new observations of radio polarization at 8\,GHz are in general
agreement with those in MBIS, although our upper limits tend to be
more conservative.  (The polarization observations quoted in MBIS were
taken from observations done in 1990--91 \cite{pat92,wil98}.) Three
sources have markedly different degrees of polarizations between the
two sets of observations: these are B\,1144+352 \& B\,1123+203
(decrease) and WLRG B\,1241+735 (increase). These differences are
commented on before discussing the sample as a whole.

The peculiar source B\,1144+352, as has already been noted, is known as
a GPS source \cite{sne95} and VLBI maps show a $\sim$ 50\,pc double
structure \cite{hen95} with components superluminally expanding
\cite{gio99}, and known to be variable (see Schoenmakers et al. 1999).
Schoenmakers et al.  (1999) argue convincingly that this is the core of
a mega-parsec sized giant radio galaxy.  The radio polarization drop may
be real, as the source was apparently near its maximum flux density at
the time of the first observations (Schoenmakers et al. Fig. 7).
Alternatively the 2$\sigma$ detection may have been erroneous,
associated with a component other than the core, or the lower resolution
of the observations presented here results in beam
depolarization.\nocite{sch99}

The candidate BL\,Lac B\,1123+203 (PGC 035156) has shown extreme
optical variability over a period of two decades, dominated by
long-term changes \cite{1988AJ.....96.1215P}, and has apparently also
suffered a polarization drop in the 6 years since the previous 8\,GHz
measurements.


\begin{figure}
\psfig{file=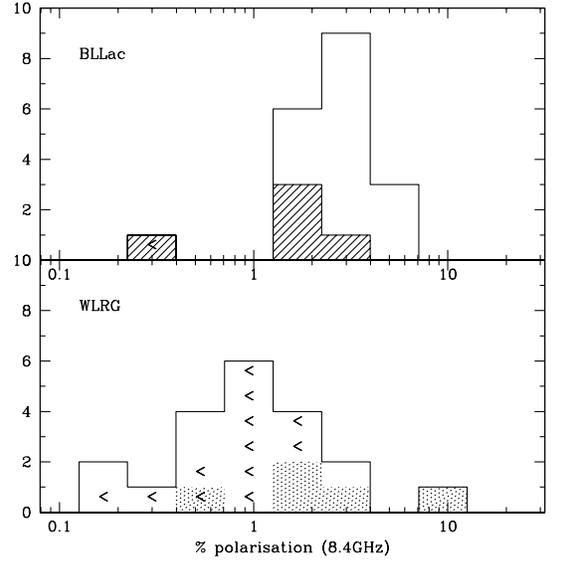,height=8cm}
\caption{Radio polarization at 8.4\,GHz, as a function of spectral class.}
\label{fig:polhist}
\end{figure}

\begin{figure}
\psfig{file=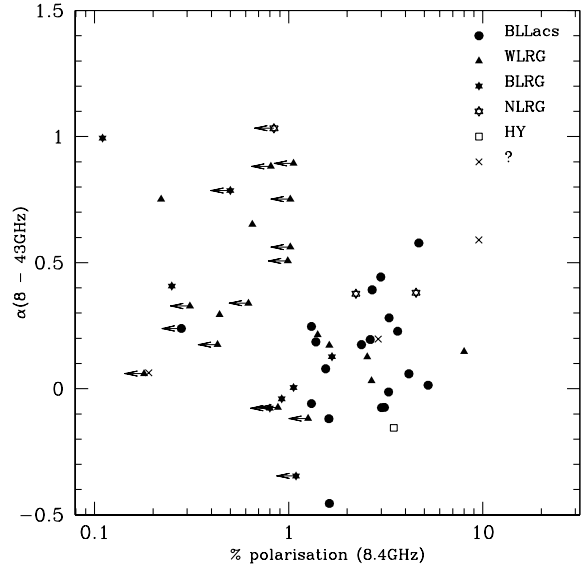,height=8cm}
\caption{Radio polarization at 8.4\,GHz, and high frequency radio spectral
index. Crosses refer to objects of unknown spectral type. Arrows are
upper limits set at {$2.5\times(\sigma Q+\sigma U)$, where $\sigma Q$
and $\sigma U$ are the noise on the Stokes Q \& U maps respectively.}}
\label{fig:pol-hi}
\end{figure}

B\,1241+735 shows very high radio polarization in these observations:
a remarkable 9\% for an object classified as a WLRG. Augusto et al. \cite*{aug98}
show this source to have a core-jet structure in  MERLIN observations.

Another source with high percentage of radio core polarization is the
`red stellar' (MBIS) object of unknown spectral type B\,2214+201. This
unusual object was the most polarized object at both epochs and may
be a star, perhaps with an extragalactic radio source along the
same line of sight.

In general we reproduce the radio polarization results of MBIS: the BL\,Lacs are
significantly more polarized at 8\,GHz than the WLRG
(Fig.~\ref{fig:polhist}).  Based on their percentage polarization at
8.4\,GHz, the probability that BL\,Lacs and WLRG are drawn from the same
population is rejected at 99.98\% level (using upper limits as
detections). It can also be seen in Fig.~\ref{fig:polhist} that the
candidate BL\,Lacs have a lower polarization than the known BL\,Lacs. The K-S
test on the BL\,Lacs and candidates shows only a 2\% chance that they are
drawn from the same population. Even with the exclusion of the source
B\,1123+203 which showed little polarization in these observations, the
populations remain distinct, with only a 6\% chance that they are from
the same population.

We also reveal a new effect: the weak emission line objects with radio
polarization $>$ 1\% also have flat high frequency radio
spectra. Fig.~\ref{fig:pol-hi} shows the 8\,GHz polarization and high
frequency spectral index for all the objects. The known WLRG in this
group of flat spectrum, high polarization sources are: B\,0149+710,
B\,0848+686, B\,1241+735, B\,1755+626 \& B\,2320+203.

\nocite{sto91}

\subsection{Variability}

 Clear evidence of variability is seen in a number of the BL\,Lacs
and BL\,Lac candidates. Studies of radio variability with the
available data of this sample are hampered by the different telescopes
used at different epochs. Nonetheless it is seen that a comparatively
large number of these sources show large differences in flux densities
between the 87GB \cite{gre91} and GB6 \cite{gre96} surveys at
4.85\,GHz, and/or the Green Bank \cite{whi92}, NVSS \cite{con98} and
FIRST \cite{whi97} surveys at 1.4\,GHz.  Indeed, the variability of
BL\,Lacs at radio wavelengths is well-known, and the subject of a
number of ongoing monitoring campaigns (see e.g. Aller et al. 1999)
\nocite{all99} which include a number of these objects.

Three other sources show evidence for variability at
1.4\,GHz as the VLA B array flux density (as measured by FIRST
 or own observations (forthcoming paper)) is greater than
the VLA D array flux density from NVSS. The VLA B array
is likely to resolve out source structure, and so we could explain a
lower B array flux density without recourse to variability, but not a
higher B array flux density. The non-BL\,Lac sources which fall in this
latter category are: two WLRGs B\,0729+562, B\,2202+363 and one object
of unknown spectral type B\,2319+317.


\section{Discussion}
\label{sect:discussion}

\subsection{The status of the `candidate' BL\,Lacs}

As the objects classified as `candidate BL\,Lacs' occupy a different
region on the contrast-EW plane than the other BL\,Lacs, we first
address whether the data are consistent with them being the same type
of object: that is, are they otherwise identical to the other BL\,Lacs?
If they have statistically different properties, are these differences
explicable if we assume that the candidate BL\,Lacs are simply either
slightly further from our line of sight, or intrinsically weaker
(relative to their hosts)? In either of these cases the hypothesis of
MBIS that they are starlight diluted BL\,Lacs then holds.

It was shown in Sect. \ref{sect:results} that the spectral indices in
the 8--43\,GHz regime were indistinguishable between the BL\,Lacs and
the candidate BL\,Lacs. It was also shown that there was a marginal
difference in the radio--optical non-thermal spectral indices, and
that the candidate BL\,Lacs were less polarized at 8.4\,GHz. The first
piece of evidence is clearly consistent with the notion of starlight
diluted BL\,Lacs, particularly striking as they differed from the
population of WLRG. However, the remaining two differences require
some consideration, which is given below.

First we consider the marginal difference in $\alpha^{\rm rad}_{\rm opt*}$. In a
sample such as this which is selected with two flux density limits
(i.e. both in the radio and in the optical), we must consider the
effects of this selection on the properties of our sample. The optical
magnitude limit implies that for objects which are dominated by
starlight there will be a smaller flux density attributable to the
non-thermal emission at 5500\,\AA: as there is no difference in the
total optical emission from the two classes of objects, we expect that
the optical emission due to the non-thermal AGN is weaker in the
objects classified as candidate BL\,Lacs. Indeed, this is so, but the
genuine BL\,Lacs are also brighter in the radio, which makes it unclear
as to whether the difference in spectra is due solely to this
selection effect. Therefore no firm conclusions can
be based upon the marginal difference: the difference could arise
simply due to selection effects, or may reflect a real difference, or
simply small number statistics (there are only 4 candidate BL\,Lacs with
calculated $\alpha^{\rm rad}_{\rm opt*}$).

The lower fractional polarization of the candidate BL\,Lacs cannot,
however, be attributed to a selection effect. Simple theories of BL\,Lac
emission with a synchrotron origin for both the radio and the optical
emission could be used to deduce constraints on the intrinsic conditions
or angle to the line-of-sight in these objects, but small number
statistics constrain us to discuss this only in qualitative terms. One
possibility is that the candidate BL\,Lacs have lower radio polarization
because this sample is at slightly larger angle to the line of sight
than the other BL\,Lacs. In this way they are objects which are more
`starlight diluted' principally because they are less Doppler boosted
due to a line of sight effect, rather than due to the relative intrinsic
weakness of the source.

\begin{figure}
\psfig{file=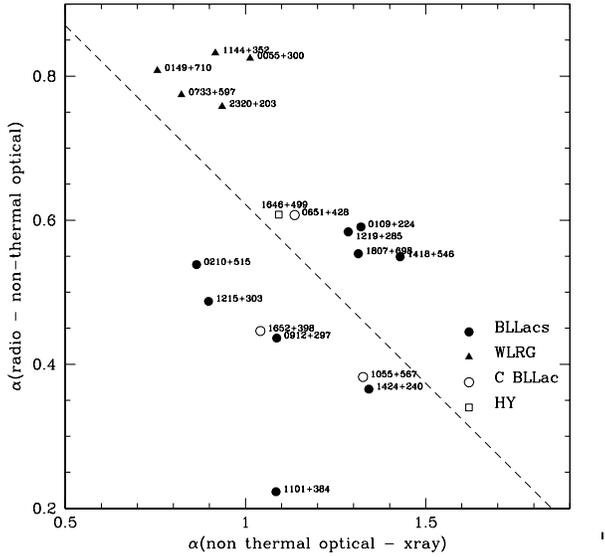,height=8cm}
\caption{The radio-optical-xray of the objects in the sample. The
Radio flux densities are those tabulated in Table~\ref{tab:res}, except
for B\,1646+499 \& B\,1652+398 for which we use the 8.4\,GHz flux
densities of Patnaik et al. (1992). The X-ray flux densities are
from the ROSAT All-Sky Survey. Only sources detected in this survey
are plotted. We plot BL\,Lacs and candidates (CBL\,Lacs) as well as the
WLRG and the detected hybrid B\,1646+499. The dashed line corresponds
to the `dividing line' between HBLs and LBLs (after Padovani \& Giommi
1995) }
\label{fig:rox}
\end{figure}
\nocite{pat92}

One potential problem with this idea is that, according to the
`accelerating jet' model (Marscher 1980; Ghisellini \& Maraschi 1989)
for objects dominated by this accelerating region of the jet, the
spectrum typically steepens as objects move into the line of
sight. This is the opposite of what, if anything, is observed as we
move from the BL\,Lacs to the candidates, although as we have pointed
out, there is a problem of a selection effect here.  
\nocite{mar80,ghi89}

On the other hand, intrinsically weaker objects may have different
synchrotron turn-over frequencies (or maximum particle
energies). Ghisellini (1997) suggests
that the difference in blazar spectral energy distributions could be
explicable by differing amounts of radiative cooling due to
electrons. These electrons could be from the BLR. In this scheme the
candidate BL\,Lacs, with strong\-er emission lines might suffer stronger
radiative cooling, and therefore be expected to have a steeper
$\alpha^{\rm rad}_{\rm opt*}$. This could only explain the difference in
polarization properties if this was due to depolarization in a medium
associated with the emission line region.  \nocite{ghi97}

The discussion here has parallels to the debate over the relation
between X-ray-- and radio--selected BL\,Lacs (XBLs \& RBLs) or,
complimentarily, high-- and low--energy cut-off BL\,Lacs (HBLs \&
LBLs) (see e.g. Maraschi et al. 1986; Padovani \& Giommi 1995): are
the differences intrinsic or due to a line of sight difference? The
accelerating jet model has had a number of difficulties in reproducing
the characteristics of the XBL and RBL populations, and recent work
has focussed either on intrinsic physical differences (Padovani \&
Giommi 1995) or on a combination of both (Georganopoulos \& Marscher
1998). Indeed these parallel discussions may be no coincidence -- it
could be that our candidate BL\,Lacs are `XBL-like': Mrk501 has been
called a 'candidate' by us, and was called XBL-like by Padovani \&
Giommi (1995).  In this light we plot (Fig.~\ref{fig:rox}) the
$\alpha_{\rm o}^{\rm r} - \alpha^{\rm o}_{\rm x}$ plane, which has
divided these populations until the more recent discovery of an
`intermediate' population \cite{per98,lau99}.  We plot the non-thermal
contribution to the optical flux density only (the plots in the
literature use the entire optical flux density). This has the effect
of moving the points left and up, although for the BL\,Lacs this
effect is negligible.  By comparison with Stocke et al. (1985) we see
that all our BL\,Lacs fall somewhat nearer the `dividing line' (illustrated
Fig.~\ref{fig:rox}) than either the complete 1\,Jy and EMSS samples, which lie
to the top right and bottom left respectively. Furthermore, the
candidates, as well as the other BL\,Lacs, appear both RBL-like and
XBL-like by this measure.  \nocite{pad95,geo98,mar86,sto85}

Our spectral results for the BL\,Lac population as a whole are
consistent with those of Gear et al. (1994), who find for a subset of
the Stickel et al. (1991) sample of BL\,Lacs, flat or rising spectra in
the range 5 to 37\,GHz ($\alpha = -0.04\pm0.20$), and falling spectra
in the range 150 to 375\,GHz ($\alpha = 0.48\pm 0.22$).  They also
find, for the cases they have data, that the 90\,GHz points are
consistent with a single power law from 90\,GHz to the millimeter
regime.  Bloom et al. (1994) also find falling spectra in the
millimeter and sub-millimeter regime.  These studies concluded that the
spectral break for radio-selected BL\,Lac objects must lie in the range
10 to 100\,GHz.  For our sample we find flat or slightly falling
spectra, with some examples of rising spectra in the range 8 to
43\,GHz.  Our observations are generally well fit by a single
power-law, implying that for our objects the break must lie $>$
22\,GHz (The 43\,GHz points not having high enough signal to noise to
rule out the start of a break in the spectra), or that the spectral
break is not sharp (so would be undetected over such a small frequency
range).  The simplest interpretation is that this sample of somewhat
fainter radio-selected BL\,Lacs has a gradual spectral turn-over at
around the same frequencies (10-100\,GHz) as the more powerful Stickel
et al. (`1\,Jy') sample.\nocite{sti91,gea94,blo94}

We therefore conclude that the results can be explained by the
hypothesis that the candidate BL\,Lacs are starlight diluted BL\,Lacs. The
fact that the candidate BL\,Lacs are indistinguishable from the BL\,Lacs
in the high frequency radio regime, with flat spectra extending out to
43\,GHz, indicates that the candidate BL\,Lacs are still highly Doppler
boosted.  If the lower radio polarization of the candidate BL\,Lacs is
explained by requiring the candidate BL\,Lacs to be slightly further
from the line of sight than the other BL\,Lacs, the marginal difference
in $\alpha^{\rm rad}_{\rm opt*}$ may reflect a selection effect.

\subsection{The relationship of the WLRG and BL\,Lacs}
\label{sect:hidd}

We wish to address whether the WLRG are consistent with being the
parent population of the BL\,Lacs: that is BL\,Lacs away from the line of
sight, and, further, whether there is any indication that any of the
objects in this class are `hidden' BL\,Lacs: that is they are oriented
at similar angles as the BL\,Lacs, but that the starlight contribution
from their hosts swamps the characteristic non-thermal emission at
optical wavelengths. 

Unlike the situation for the BL\,Lacs themselves we cannot assume that
the detected emission of the WLRG comes from jet-like structures.
Augusto et al (1998) presented VLBI observations which contained a
number of the sources in this sample, and find the WLRGs B\,0116+319
to be a CSO, and B\,1241+735 to be a core--jet.

The mean 8--43\,GHz spectral index is steeper for the WLRG than the
BL\,Lacs. This is consistent with both emission from more extended
regions and jet emission at larger angles to the line of sight, if the
jet is no longer dominated by an accelerating inner jet. In the
latter case this arises as the break in the spectrum (in BL\,Lacs at a
few tens of GHz) is not Doppler shifted to such high frequencies.

At centimeter wavelengths with the resolution of the observations
presented here, detailed interpretation of the polarization properties
of the sample are not possible, as the effects of Faraday rotation and
depolarization and jet bending complicate the interpretation. However
the results are consistent with the canonical jet-in-shock model
(e.g. Blandford \& Konigl, 1979)\nocite{bla79} If we consider the
emission detected in the WLRG to be jet material, the comparatively
unpolarized cores of the WLRG in comparison to the BL\,Lacs could
arise either by Faraday depolarization by a surrounding dense
magneto-ionic medium (e.g. torus) or by the dominance of a single
Doppler--boosted shock structure in the BL\,Lacs.  If the polarized
emission comes from a variety of structures, or from the general jet
material, we must favour Faraday depolarization of the WLRG to explain
the difference in polarization properties. A further possibility is
that the emission detected in the WLRG is not parsec-scale jet
material as is it is for the BL\,Lacs: indeed, as mentioned above, the
WLRG contain a number of known compact symmetric objects. The emission
from these objects come from a variety of structures, including
hotspots and any lobes, and in these cases the lack of polarization is
most likely to be caused by beam depolarization.  \nocite{mar80}

Using the Rosat All-Sky Survey and the data presented in this paper,
we suggest the following 5 sources contain AGN very close to the line of
sight:\\ 
{\bf B\,0149+710} has a RASS detection, high \%P$_{\rm radio}$ and flat
$\alpha^{8}_{43}$; it has a relatively low Ca\,II break
contrast, and was considered a candidate BL\,Lac in MBIS (see
Sect.~\ref{sec:opt})\\
{\bf B\,1144+352} too was considered a candidate \\BL\,Lac in
MBIS, and has a RASS detection, as well as observed superluminal motion (Giovannini et al
1999); \\ 
{\bf B\,1241+735} is a core-jet source (Augusto et al. 1998) has a very
high  \%P$_{\rm radio}$ and flat $\alpha^{8}_{43}$\\
{\bf B\,1755+626} has a high \%P$_{\rm radio}$ and flat $\alpha^{8}_{43}$\\
{\bf B\,2320+203} has a  RASS detection and a high \%P$_{\rm radio}$ and flat
$\alpha^{8}_{43}$\\
These sources are potentially `hidden' BL\,Lacs: those which are not
detected by the measurement of the Ca\,II break, because their
non-thermal continuum is swamped by starlight. These sources are shaded
in the lower panels of Figs. 2, 4 and 6. These sources have similar
$\alpha^{\rm o}_{\rm x}$ to the BL\,Lacs, but steeper radio to optical spectra.

The other two RASS detections, namely \\B\,0055+300 and B\,0733+597 are
also candidates of beamed AGN.  B\,0055+300 (NGC315) is a giant FRI,
with an asymmetric two-sided jet and prominent core (the reason 
the source was included in the 200mJy sample). Due to the core
brightness Venturi et al. (1993) argued the jet was highly
relativistic. VLBI observations (Cotton et al. 1999) indicate Doppler
favoritism as the cause of the brightness asymmetry on small scale and
argue the jet is aligned at 35$^\circ$ to the line of
sight. B\,0733+597 shows a core-jet structure with a faint counter-jet
to the south on VLBI scales (Taylor et al. 1994). \nocite{cot99,ven93,tay94}

One hybrid (B\,1646+499) was detected in the RASS, but unfortunately
we have no high frequency radio data for this source. However in MBIS
it was relatively strongly polarized at 8.4\,GHz (1\%). The other
hybrid (B\,0125+487) was also strongly polarized at 8.4\,GHz (1.9\% in
MBIS, 3.4\% here) and had a slightly rising $\alpha^{43}_8$ spectrum.


From the steep $\alpha^{8}_{43}$, we suggest that B\,0035+227,
 B\,0116+319, B\,1217+295 and B\,1558+595 are likely not to be highly
 boosted objects, and may, like B\,0116+319, turn out to be
 CSOs. B\,0729+562 is intriguing because it too has a steep
 $\alpha^{8}_{43}$, but has apparently shown variations at 1.4\,GHz.

In summary, the steeper $\alpha^8_{43}$ and lower radio polarizations
of the WLRG as a population as a whole when compared to the BL\,Lacs
can be explained if the WLRG are, on average, at larger angles to the
line of sight. The fact that WLRG with radio polarization $>$ 1\% were
found to have flatter $\alpha^8_{43}$, can therefore be interpreted as
identifying a sub-sample of WLRG closest to the line of sight.  The
high incidence of X-ray detections in this sub-sample supports this
claim. We suggest that around a quarter of the WLRG in the sample are
these `hidden' BL\,Lacs. A similar number of sources may well turn out
to be compact objects which are not highly relativistically boosted,
but whose spectra is relatively flat at 1.4--5\,GHz, as synchrotron
losses do not yet dominate the sources.  These WLRG may not be so
directly related to the BL\,Lacs in the sample, and this should be
borne in mind in further statistical studies. The remaining half of
the sources we speculate to be predominantly core-dominated FRI-like
sources, but which are not so close to the line of sight to be
considered `hidden BL\,Lacs'.

 

\section{Conclusions}

We have shown that the radio and radio--optical spectra are consistent
with the `candidate' BL\,Lacs of MBIS (revised definition in
Sect.~\ref{sec:opt} for compatibility with other workers) being
genuine \\BL\,Lacs. However, the percentage radio polarization may
indicate some slight differences between these groups. In particular:

\begin{itemize}
\item{The spectral indices at high radio frequencies indicate that the
    BL\,Lacs and candidate BL\,Lacs are indistinguishable, and that the
    sample of WLRG as a whole is statistically distinct from the BL\,Lacs
    in this regime.}
\item{The candidate BL\,Lacs do however show a \\marginally steeper
    radio--optical spectra than the known BL\,Lacs. This is consistent
    with them being starlight--diluted BL\,Lacs in a sample limited in
    both optical and radio flux density.}
\item{ The candidate BL\,Lacs show a somewhat smaller polarization at
   radio wavelengths than the known BL\,Lacs. This could indicate they
   are slightly further from the line of sight, or that (possibly) they
   have a greater Faraday depth of depolarising medium.}
\end{itemize}

The WLRG population of the 200\,mJy sample is composed primarily of
objects dominated by relativistically boosted cores and jets, which
are probably closely related to the BL\,Lacs, as well as objects which
just creep into the flat--spectrum sample with $\alpha^{1.4}_{5} <
0.5$, but whose spectra steepen more rapidly above 5\,GHz. We confirm
the result of MBIS that the BL\,Lacs have higher radio polarization
cores compared to the WLRG, consistent with a simple interpretation as
the WLRG being, on average, further from the line of sight.  We also
show that WLRG with highly polarized cores have flatter high frequency
radio spectra, and suggest that these are the WLRG closest to the line
of sight.

Within the WLRG population, we identify five objects which we believe
    to be strongly relativistically boosted. These objects are
    candidates for `hidden BL\,Lacs': i.e. those objects physically
    identical to BL\,Lacs, at similar angles to the line of sight, but
    whose intrinsic power relative to their host galaxy makes them
    undetected as BL\,Lacs in the optical regime.

\begin{acknowledgements}
We thank Greg Taylor and Michael Rupen for help with the VLA
scheduling. The National Radio Astronomy Observatory is a facility of
the National Science Foundation operated under cooperative agreement
by Associated Universities, Inc.  This research was supported by the
European Commission, TMR Programmme, Research Network Contract
ERBFMRXCT96-0034 `CERES'.

\end{acknowledgements}

\bibliographystyle{aabib99}

\end{document}